\documentclass[12pt]{article}

\newcommand{\be}{\begin{equation}}
\newcommand{\ee}{\end{equation}}
\newcommand{\beq}{\begin{eqnarray}}
\newcommand{\eeq}{\end{eqnarray}}
\newcommand{\crs}{cross section}
\newcommand{\crss}{\crs s}
\newcommand{\PLA}{p_{_{\rm LAB}}\!}

\newcommand{\hf}{\hat f}
\newcommand{\hff}{\hat F}
\newcommand{\hmm}{\hat M}
\newcommand{\dist}{\displaystyle}
\newcommand{\snn}{\sqrt{s}^{}_{N\!N}}
\newcommand{\skn}{\sqrt{s}^{}_{K\!N_i}}
\newcommand{\rpt}{\rule{0pt}{14pt}}

\newcommand{\rppt}{\rule{0pt}{22pt}}
\newcommand{\rpttt}{\rule{0pt}{30pt}}
\newcommand{\dso}{d\sigma /d\Omega}
\newcommand{\half}{\frac{1}{2}}
\newcommand{\bftau}{\mbox {\boldmath $\tau$}}

\newcommand{\bfKN}{\mbox {\boldmath $K\!N$}}
\newcommand{\bfKD}{\mbox {\boldmath $Kd$}}
\newcommand{\bfKpD}{\mbox {\boldmath $K^+d$}}
\newcommand{\bfKpDD}{\mbox {\boldmath $\,K^{+\!}D\,$}}
\newcommand{\bfKDNN}{\mbox {\boldmath $Kd\!\to\!K\!N\!N$}}
\newcommand{\bfKNI}{\mbox {\boldmath $K\!N(I\!=\!0)$}}
\newcommand{\KDNN}{Kd\!\to\!K\!N\!N}
\newcommand{\kdpn}{K^+d\!\to\!K^+ pn}
\newcommand{\kdpp}{K^+d\!\to\!K^0 pp}
\newcommand{\kdkd}{K^+d\!\to\!K^+ d}
\newcommand{\phase}{\delta^{I}_{l\pm}}
\newcommand{\stot}{\sigma^{tot}_{K^+d}}
\newcommand{\sela}{\sigma^{el}_{K^+d}}
\newcommand{\sine}{\sigma^{inel}_{K^+d}}

\newcommand{\bfp}{\mbox {\boldmath $p$}}
\newcommand{\bfq}{\mbox {\boldmath $q$}}
\newcommand{\bfpp}{\mbox {\boldmath $P$}}
\newcommand{\bfqq}{\mbox {\boldmath $Q$}}
\newcommand{\hPsi}{\hat\Psi}
\newcommand{\hPhi}{\hat\Phi}
\newcommand{\bfsig}{\mbox {\boldmath $\sigma$}}
\newcommand{\bfeps}{\mbox {\boldmath $\epsilon$}}
\newcommand{\bfdeL}{\mbox {\boldmath $\Delta$}}
\newcommand{\bfn}{\mbox {\boldmath $n$}}
\newcommand{\bfk}{\mbox {\boldmath $k$}}

\newcommand{\veps}{\varepsilon}
\newcommand{\inpppp}{\int\!\frac{d^3\bfp_4}{(2\pi)^3}}
\newcommand{\intdp}{\int\!\frac{d^3\bfp}{(2\pi)^3}}
\newcommand{\intdq}{\int\!\frac{d^3\bfq}{(2\pi)^3}}
\newcommand{\bfr}{\mbox {\boldmath $r$}}
\newcommand{\dij}{\delta^{}_{ij}}

\newcommand{\abra}{\left(\!\! \begin{array}{cc} }
\newcommand{\aket}{\end{array}\!\!\right)}
\date{}

\title{\bf On the \bfKpDD\ interaction at low energies}
\author{V.~E.~Tarasov, V.~V.~Khabarov,
        A.~E.~Kudryavtsev, V.~M.~Weinberg \\
  \it Institute of Theoretical and Experimental Physics \\
  \it Moscow, Russia}

\begin{document}
\maketitle

\vspace{-10mm}

\begin{abstract}
The $Kd$ reactions are considered in the impulse approximation with
NN final-state interactions (NN~FSI) taken into account. The realistic
parameters for the $KN$ phase shifts are used. The "quasi-elastic"
energy region, in which the elementary $KN$ interaction is predominantly
elastic, is considered. The theoretical predictions are compared with
the data on the $\kdpn$, $\kdpp$, $\kdkd$ and $K^+d$ total \crss.
The NN~FSI effect in the reaction $\kdpn$ has been found to be large.
The predictions for the $Kd$ \crss\ are also given for slow kaons,
produced from $\phi(1020)$ decays, as the functions of the isoscalar
$KN$ scattering length $a^{}_0$. These predictions can be used to
extract the value of $a^{}_0$ from the data.

\vspace{5mm}

\end{abstract}

\section{\bf Introduction}

Recent experimental indications~\cite{penta} on the possible existence
of the exotic $\Theta^+(1540)$ state in the $K^+N$ system (pentaquark
$\bar suudd$ state) have enhanced the interest to the $K^+N$ and $K^+d$
interactions. Analyses of the $K^+N$ elastic scattering
\cite{Arndt,Haid}, $K^+d$ total \crss~\cite{Nuss,Gibbs} and the
reaction $\kdpp$~\cite{Cahn,Sib} with isospin $I=0$ assumed for the
pentaquark led to conclusion about its small width $\Gamma\le 1$~MeV.
At present time the existence of the pentaquark seems to be doubtful
since it was not confirmed by a number of experiments (see the review
paper by M.~Danilov and R.~Mizuk listed in Ref.~\cite{penta}).

In connection with the $\Theta^+$ problem the recent paper~\cite{Gibbs1}
introduces a new partial-wave analysis (PWA) of $K^+N$ scattering in
the momentum range $0<\PLA<1.5$~GeV$/c$ and comparison of the
results with previous analyses. To extract the $K^+N$-scattering
parameters in isospin-0 channel an additional information to the data
on the proton target is required. This additional data are provided by
experiments on the $K^+$-deutron collisions. The recent
paper~\cite{Sib1} contains a review of existing data on the $K^+d$
reactions in the "quasi-elastic" region $\PLA<0.8$~GeV$/c$
(where the elementary $KN$ reaction is predominantly elastic and the
role of the particle-production processes is negligible), i.e. on the
processes $K^+d\to K\!N\!N$ and $K^+d\to K^+d$. A reasonable description
of the total \crss\ and differential spectra $\dso$ for outgoing kaons
was obtained in the framework of the impulse approximation with the use
of the J\"ulich model for the $KN$ amplitude.
The papers~\cite{Gibbs1,Sib1} also claim that the existing $K^+N$ and
$K^+d$ data do not prove, but do not exclude the possibility of
a narrow resonance $\Theta^+$ in the $KN(I=0)$ system.

In this paper we present the calculations of the total $K^+d$ \crss,
the \crss\ of the break-up reactions $K^+d\to K^+pn,K^0pp$ and of the
elastic scattering process $K^+d\to K^+d$. We restrict our
consideration to the "quasi-elastic" region $\PLA<0.8$~GeV$/c$ defined
above. We consider only the "background" amplitudes and neglect the
possible $\Theta^+$ contribution. In our calculations we take into
account the pole diagrams and $s$-wave final-state interaction (NN~FSI)
of slow nucleons in the process $Kd\to K\!N\!N$. We express the
$KN$-scattering (and charge exchange) amplitude through the partial-wave
components, including only $s$- and $p$-wave terms which are important
in the region of interest.
For simplicity we use the $s$-wave $KN$-scattering amplitude when
calculating the NN~FSI contribution. In this approximation the $KN$
amplitude does not contain the spin-flip term and as a result the NN~FSI
is taken into account only in the triplet $N\!N(^3 S_1)$ state.
Thus, according to the Pauli principle, in the case under consideration
NN~FSI takes place only for the process $K^+d\to K^+pn$ but not for
$K^+d\to K^0pp$.

Predictions will be also given for the $K^+d$ \crss\ with slow kaons
as functions of the isoscalar $KN$-scattering length $a_0$. They can be
used to extract the value of $a_0$ from the experiments with the kaons
produced from $\phi$(1020) decays at rest. The corresponding
experiments with $\phi$(1020) mesons produced in $e^+e^-$-collisions
may be proposed for DA$\Phi$NE machine (Frascatti).

The paper is organized as follows. In Sect.~2 we give the expressions
for the partial-wave $KN$-scattering amplitudes and illustrate the
description of the data on the $KN$ \crss\ for some sets of phase-shift
parameters. In Sect.~3 we write out the amplitudes for the break-up
reactions $Kd\to K\!N\!N$ (pole diagrams $+$ NN~FSI) and the elastic
process $K^+d\!\to\!K^+d$. In Sect.~4 our theoretical predictions for
the $K^+d$ \crss\ are presented and compared with the experimental data.
Sect.~5 is the Conclusion. Some necessary formulas are placed in the
Appendix.

\section{\bf \bfKN-scattering amplitude}

Let us write out the isospin structure of the $KN$-scattering amplitude
$\hf^{}_{KN}$ and the relations between charge and isospin amplitudes.
They read
\be
\hf^{}_{K\!N}=\hff^{}_S+\hff^{}_V \bftau^{}_K\bftau,~~~
\hff^{}_S=\frac{1}{4}(3\hff_1+\hff_0),~~~~
\hff^{}_V=\frac{1}{4}(\hff_1-\hff_0),
\label{1}\ee
$$
\hf^{}_{K^+p}=\hff^{}_1,~~~~
\hf^{}_{K^+n}=\half (\hff^{}_1\!+\!\hff^{}_0),~~~~
\hf^{}_{K^+n\to K^0p}=\half(\hff^{}_1\!-\!\hff^{}_0),
$$
where $\bftau$ ($\bftau^{}_K$) are the isospin Pauli matrices for
nucleons (kaons) and $\hff^{}_{0,1}$ ($\hff^{}_{S,V}$) are the $K\!N$
amplitudes with $s$-channel ($t$-channel) isospin $I=0,1$. The
$K^+N$-scattering and charge exchange amplitudes
are\footnote{
With these amplitudes $\hf$ the differential \crs\ for binary
reaction is $\dso=|\varphi^+_2\hf\varphi^{}_1|^2 q_2/q_1$.
Here: $\varphi^{}_{1,2}$ are the spinors (and isospinors) of the
initial and final nucleon and $\varphi^+_i\varphi^{}_i\equiv\! 1$;
$q^{}_{1,2}$ are the initial and final relative momenta.
Throughout the paper we use the word "amplitude" for the matrix element
$\varphi^+_2\hf\varphi^{}_1$ and for the operator $\hf$ as well.}
\be
\hf^{}_{K^+\!N\to K^+N}\!=\!\hff^{}_S+\!\hff^{}_V\tau_{3\,},~~~
\hf^{}_{K^+\!n\to K^0 p}\!=\!2\hff^{}_V\tau_{+\,},~~~
\tau_+\!=\half(\tau_1\!+\!i\tau_2).
\label{2}\ee
The amplitudes $\hff^{}_I$ ($I=0,1$) can be written in standard form
$$
\hff_I=A_I+B_I (\bfn\bfsig),~~
\bfn=\frac{\bfk\times\bfk'}{|\bfk\times\bfk'|},~~
A_I=\sum^{\infty}_{l=0}\left[(l+1)f^{(I)}_{l+}+lf^{(I)}_{l-}\right]
P^{}_l(\cos\theta),
$$
\be
B_I=i\sum^{\infty}_{l=1}\left(f^{(I)}_{l+}-f^{(I)}_{l-}\right)
P^1_l(\cos\theta),~~~ f^{(I)}_{l\pm}=\frac{\eta\exp(2i\delta)-1}{2ik}.
\label{3}\ee
Here: $\bfk$ ($\bfk'$) is the relative initial (final) CM momentum;
$\eta\!\equiv\!\eta^{(I)}_{l\pm}$ and $\delta\!\equiv\!\phase$ are the
inelasticities ($0\!\le\!\eta\!\le\!1$) and phase shifts for
$s$-channel isospin $I$, orbital momentum $l$ and $j=l\pm\half$.
At $\PLA<0.8$~GeV$/c$ the \crss\ of particle production
in the $KN$ interactions are relatively small and
$\eta^{(I)}_{l\pm}\!\approx\! 1$. Hereafter, we take
$\eta^{(I)}_{l\pm}\!\equiv\! 1$ and
$f^{(I)}_{l\pm}\!=\!1/(k\cot(\delta)-ik)$.

Let us compare the $KN$ \crss, calculated from the existing phase
shifts $\phase$, with the data. Here we use the phase shifts from
Refs.~\cite{Gibbs,Gibbs1} given in the forms
\be
k\,\cot(\phase)=\frac{1}{a^{(I)}_{l\pm} k^{2l}}~~~
(l=0,1)~\cite{Gibbs},
\label{4}\ee
\be
k\,\cot(\phase)=\frac{1}{a^{(I)}_{l\pm} k^{2l}}\,
\Bigl(\,1+\sum_{i\ge 1} b^{(I)}_{i,l\pm} k^{2i}\,\Bigr)~~
(l=0,1)~\cite{Gibbs1},
\label{41}\ee
$$
\delta^{(I)}_{2\pm}=k^5_{} (c^{(I)}_{\pm}+d^{(I)}_{\pm} k^2
+e^{(I)}_{\pm} k^4),~~
\delta^{(I)}_{3\pm}=k^7_{} f^{(I)}_{\pm}~\cite{Gibbs1},
$$
where $a^{(I)}_{l\pm}$, $b^{(I)}_{i,l\pm}$, $c^{(I)}_{\pm}$,
$d^{(I)}_{\pm}$, $e^{(I)}_{\pm}$, and $f^{(I)}_{\pm}$ are the constants.
Their values were obtained~\cite{Gibbs1} from PWA of $K^+N$ scattering
in the range $\PLA<1.5$~GeV$/c$, using $s$-, $p$-, $d$-, and
$f$-wave amplitudes (see Tables~II and VI in Ref.~\cite{Gibbs1}). The
values $a^{(I)}_{l\pm}$ for $l=0,1$ (scattering lengths and volumes)
from Refs.~\cite{Gibbs,Gibbs1} are given in Table.

Fig.~1 shows the total $K^+p$, $K^+n$ and $K^+N(I\!=\!0)$ \crss.
The symbols represent the experimental data, taken from
Refs.~\cite{Bow70,Bow73,Bug68,Car73,Burn74,Kycia,Adams71,Adams73,
Came74,Cook61,Cool70,Gia72}. Fig.~1$a$ shows the data on the total
($\sigma^t_{K^+p}$) and elastic ($\sigma^{el}_{K^+p}$) $K^+p$ \crss.
The curves in Figs.~1$a$ and 1$b$ show the calculated values
 $\sigma^t_{K^+p}=\sigma^{el}_{K^+p}$ ($a$) and
 $\sigma^t_{K^+n}=\sigma^{el}_{K^+n}+\sigma^{}_{K^+n\to K^0p}$ ($b$).
The \crss\ $\sigma^{I=0}_{KN}$ in Fig.~1$c$ were calculated through the
relation $\sigma^{I=0}_{KN}=2\sigma^t_{K^+n}-\sigma^t_{K^+p}$ in
accordance with the data from Refs.~\cite{Bow70,Bow73,Car73}.

The solid and dotted curves in Fig.~1 correspond to the results obtained
with the use of parameters from Ref.~\cite{Gibbs1}. The solid and dotted
curves show the contributions of the partial waves with $0\le l\le 3$
and $0\le l\le 1$, respectively. As one can see from Fig.~1 the $d$- and
$f$-wave contributions are negligible at $\PLA< 0.7$~GeV$/c$. The dashed
curves show the results obtained with the phase shifts from
Ref.~\cite{Gibbs}, where only $s$ and $p$ waves were included. Below,
calculating the $Kd$ cross sections, we use the $KN$ parameters from
Ref.~\cite{Gibbs1}.

\begin{center}
TABLE: $s$-wave scattering lengths $a_I\equiv a^{(I)}_0$ and \\
~~~$p$-wave scattering volumes $v^{\pm}_I\equiv a^{(I)}_{1\pm}$
from Refs.~\cite{Gibbs,Gibbs1}.

\vspace{2mm}
\begin{tabular}{|c|c|c|c|c|c|c|c|c|}
\hline\rpt  Ref. &
 $a_1$~(fm) & $v^-_1$~(fm$^3$) & $v^+_1$~(fm$^3$) &
 $a_0$~(fm) & $v^-_0$~(fm$^3$) & $v^+_0$~(fm$^3$) \\
\hline\rpt \cite{Gibbs} &
 -0.328 & -0.02 & 0.015 & -0.06 & 0.123 & -0.010 \\
\hline\rpt \cite{Gibbs1} &
 -0.308 & -0.092 & 0.103 & -0.1048 & 0.183 & -0.029 \\
\hline
\end{tabular} \end{center}

\section{\bf Amplitudes of the \bfKD\ reactions}

\subsection{\bf The pole amplitudes of the break-up reactions}

The pole diagrams for the $\KDNN$ reactions are shown in Fig.~2$a$,
where $M^{}_1$ and $M^{}_2$ also stand for the corresponding invariant
amplitudes (the particle momenta are given in the deuteron rest frame).
Hereafter we consider the deuteron and nucleons as nonrelativistic
particles and kaons as relativistic ones. To calculate the $Kd$
amplitudes, we use the deuteron wave function (DWF) $\Psi(\bfq)$
in the form
\be
\Psi(\bfq)=\frac{1}{\sqrt{2}}\,\varphi^+_{2\,}\hPsi(\bfq)\varphi^c_1,
~~
\hPsi(\bfq)=\frac{u(q)}{\sqrt{2}}^{\,}(\bfeps\bfsig) -\!
\frac{w(q)}{2}\left[\frac{3(\bfq\bfeps)^{\,}(\bfq\bfsig)}{q^2}
-\!(\bfeps\bfsig)\right]
\label{5}\ee
($q\!=\!|\bfq|$).
Here: $\,\varphi^{}_{1,2}$ are the nucleon spinor-isospinors
($\varphi^+_i\varphi^{}_i\equiv\! 1$); the notation $\,\varphi^c\,$
means the charge-conjugated spinor-isospinor
$\,\varphi^c\equiv\tau^{}_2\sigma^{}_2\varphi^*$; $\bfq$ and $\bfeps$
are the relative momentum of the nucleons and polarization vector of
the deuteron, respectively; $\,u(q)$ and $w(q)$ are the $s$-
and $d$-wave parts of the DWF, respectively, normalized as
$\int\!d^3q\, [u^2(q)\!+\!w^2(q)]=(2\pi)^3$.

The invariant amplitudes $M^{}_1$ and $M^{}_2$ with DWF~(\ref{5})
read\footnote{
When writing out the amplitudes of the reactions on the
deuteron we follow the diagrammatic technique of Ref.~\cite{Tar}.}
\be
M_1\!=2\sqrt{m}\,\varphi^+_1\hmm^{}_{K\!N_1}\!\hPsi(\bfp_2)\varphi^c_2,
~~~~~M_2\!=
 -2\sqrt{m}\,\varphi^+_2\hmm^{}_{K\!N_2}\!\hPsi(\bfp_1)\varphi^c_1.
\label{6}\ee
Here: $m$ is the nucleon mass; $\,\varphi^{}_i$ is the spinor-isospinor
of the $i$-th final nucleon;
$\hmm^{}_{KN_i}=8\pi\skn\,\hf^{}_{KN_i}$ is the $KN$-scattering
(on the $i$-th nucleon) invariant amplitude; $\,\skn$ is the invariant
mass of the $KN_i$ system. The \crss, expressed through the invariant
amplitudes, are given in Subsect.~1 of Appendix. The amplitude $M^{}_2$
in Eqs.~(\ref{6}) can be obtained from $M^{}_1$ by interchanging
the nucleons, i.e. $M^{}_2=-M^{}_1(N^{}_1\!\leftrightarrow \! N^{}_2)$.
Thus, the amplitude $M_1\!+\!M_2$ is antisymmetric with respect to
nucleons transposition in accordance with the Pauli principle. Further,
calculating the interference of the amplitudes $M_1$ and $M_2$, it is
convenient to rewrite one of the them, say $M_2$, with the help of
 identity  $\varphi^+_2 \hat A\varphi^c_1\equiv \varphi^+_1 \hat A^c
 \varphi^c_2$, where $\hat A$ is an arbitrary operator, containing the
Pauli spin ($\bfsig$) and isospin ($\bftau$) matrices, and
$A^c\equiv\sigma^{}_2\tau^{}_2 A^T \sigma^{}_2\tau^{}_2$ (note that
$I^c\!=\!I$, $\bfsig^c\!=\!-\bfsig$ and $\bftau^c\!=\!-\bftau$). Then
we can rewrite the amplitudes~(\ref{6}) as
\be
M_1\!=c^{}_1\,\varphi^+_1\hf^{}_{K\!N_1}\!\hPsi(\bfp_2)\varphi^c_2,~~~
M_2\!=-c^{}_2\,\varphi^+_1\hPsi^c(\bfp_1)\hf^c_{K\!N_2}\varphi^c_2,
\label{7}\ee
where $c^{}_i=16\pi\sqrt{m^{\,} s^{\phantom{x}}_{K\!N_i}}$. The
expressions for the particle momenta used to calculate the amplitudes
are given in Subsect.~2 of Appendix. The squares and interferences
of the amplitudes for the reactions $\KDNN$ with unpolarized particles
are given in Subsect.~3 of Appendix
\footnote{
There are relations, derived in Ref.~\cite{Ste64}, for the
differential \crss\ of reactions $\KDNN$ in the impulse
approximation with spin and isospin variables taken into account.
However, they can not be applyed when $d$-wave part of WFD or the
rescatterings are included.}.

Strictly speaking, in the case of the reaction $\kdpn$, the pole
amplitudes $M_{1,2}$~(\ref{7}) should also contain the term
proportional to the Coulomb $K^+p$-scattering amplitude $f^{}_C$.
It can be included by
the replacement   $\hf^{}_{K\!N_i}\!\to\!\hf^{}_{K\!N_i}\!
 +\!\half f^{}_C (1\!+\!\tau_3)$ in the hadronic $K^+\!N$-scattering
amplitudes. Since $f^{}_C\sim 1/t$, where $t\!=\!(k-k_1)^2$ is the
square of the four-momentum transfer, the Coulomb effects may be
essential at small scattering angles of the outgoing kaons (small $t$).
Thus, the measured \crss\ of the reaction $\kdpn$ should depend on
the experimental conditions.

Note that in the break-up reaction $\KDNN$ we always have
$t\le -|t|^{}_{min}<0$, since $\sqrt{s}^{}_{N\!N}\ne m_d$ ($m_d$,
$\sqrt{s}^{}_{N\!N}$ are the deuteron mass and the effective mass of
the $N\!N$ system). Thus, the total Coulomb \crs\ $\sigma^{}_C(\kdpn)$
is finite unlike the case of the elastic scattering processes for which
it diverges at zero scattering angles (at $t\!\to\!0$). An estimate
of the Coulomb \crs\ $\sigma^{}_C(\kdpn)$ is given in Subsect.~6
of Appendix. In the following we neglect the Coulomb effects when
calculating the $Kd$ \crss.

\subsection{\bf The final-state interaction (FSI) of nucleons}

At low energies the important effect should come from the
nucleon-nucleon final state interaction (FSI) due to large
$NN$-scattering lengths. The famous FSI effect is responsible for
the near-threshold enhancement in the mass spectra $d\sigma/d\snn$
(Migdal-Watson effect~\cite{Mig,Wat}) of the meson production reactions
$N\!N\!\to\!N\!N x$ and increases the reaction \crs\ in the
near-threshold region. The role of secondary rescatterings was also
investigated in reactions on the deuteron (for example, in the reaction
$\pi^-d\!\to\!\pi^-pn$~\cite{KK}). In this paper we take into account
only the $NN$-rescattering amplitudes and neglect the
$KN$-rescatterings, since the $KN$-scattering lengths are relatively
small, $a^{}_{K\!N}\ll a^{}_{N\!N}$.

\vspace{1mm}
The NN~FSI diagram is shown on Fig.~2$b$ (the diagram $M^{}_R$). We
consider only the $s$-wave $NN$ rescattering. It is convenient to
write the invariant amplitude of the $s$-wave scattering
$N^{}_3N^{}_4\!\to\! N^{}_1N^{}_2$ in the form
\be
M^{(S)}_{N\!N}= 8\pi \snn\, \Bigl[f^{(0)}_{N\!N}(p)^{\,}
   (\varphi^{c+}_4\frac{\bfsig}{\sqrt{2}}\varphi_3)^{\,}
   (\varphi^+_1\frac{\bfsig}{\sqrt{2}}\varphi^c_2)
  +\,f^{(1)}_{N\!N}(p)^{\,}
(\varphi^{c+}_4\frac{\bftau}{\sqrt{2}}\varphi_3)^{\,}
(\varphi^+_1\frac{\bftau}{\sqrt{2}}\varphi^c_2)\Bigr],
\label{9}\ee
where $\snn$ is the invariant mass of the $NN$ system (we take
$\snn\!=\!2m$ for the nonrelativistic nucleons) and
$f^{(I=0,1)}_{N\!N}(p)$ are the $NN$-scattering amplitudes with isospin
$I$ and relative momentum $p$, normalized as $\dso=|f|^2$. The first
(second) term in Eq.~(\ref{9}) corresponds to the $NN$-scattering
amplitude with isospin $I=0(1)$ and total spin $S^{}_{NN}=1(0)$
in accordance with the Pauli principle.

Let the $N_3N_4\!\to\! N_1N_2$ amplitude be given in the general form
 $M^{}_{N\!N}=(\varphi^{c+}_4\hat B_{2\,}\varphi_3)^{\,}
(\varphi^+_1\hat B_{1\,}\varphi^c_2)$, where the operators
$\hat B^{}_{1,2}$ contain the nucleon spin ($\sigma$) and isospin
($\tau$) Pauli matrices. Then, making use of the notations
$\hPsi(\bfp^{}_i)$ and $\hmm^{}_{K\!N}$ from Eqs.~(\ref{5}) and
(\ref{6}), one can obtain the NN~FSI amplitude $M^{}_R$ in the form
\be
M^{}_R=\frac{-1}{\sqrt{m}}\inpppp\,
\frac{{\rm Tr}\{\hPsi(\bfp_4)\hat B_2\hmm^{}_{K\!N}\} }
{2m\veps_3-\bfp^2_3+i0}\,(\varphi^+_1\hat B_1\varphi^c_2).
\label{10}\ee
Here: $\!\bfp^{}_i$ and $\veps^{}_i$ ($i\!=\!1,2,3,4$) are the nucleon
momenta in the deutron rest frame (see Fig.~2$b$) and kinetic energies,
respectively; the nucleon with the momentum $\bfp^{}_4$ is "on-shell";
$\veps_3=\veps_1+\veps_2-\veps_4$, $\veps^{}_{1,2,4\!}=p^2_{1,2,4}/2m$;
Tr$\{(\cdots)\}\equiv {\rm Tr}\{\hat T\}^{\,}{\rm Tr}\{\hat S\}$,
where Tr$\{\hat T\}$ (Tr$\{\hat S\}$) is a trace of isospin (spin) part
$\hat T$ ($\hat S$) of the matrix expression $(\cdots)=\hat T\hat S$.
Taking the $NN$ amplitude, given by Eq.~(\ref{9}), we should make the
replacement
\be
\begin{array}{ll}
\!{\rm Tr}\{\hPsi\hat B_2\hmm^{}_{K\!N}\}
(\varphi^+_1\hat B_1\varphi^c_2)\to\!\!  &
 8\pi m\,\Bigl[ f^{(0)\,off}_{N\!N}\!(q,p)^{\,}
{\rm Tr}\{\hPsi\,\bfsig\hmm^{}_{K\!N}\}(\varphi^+_1\bfsig\varphi^c_2)\\
\rule{0pt}{24pt}    &  ~~+ f^{(1)\,off}_{N\!N}\!(q,p)^{\,}
{\rm Tr}\{\hPsi\,\bftau\hmm^{}_{K\!N}\}
(\varphi^+_1\bftau\varphi^c_2)\Bigr]
\end{array}
\label{101}\ee
in Eq.~(\ref{10}). Here $f^{(I)\,off}_{N\!N}(q,p)$ is the "off-shell"
$NN$ amplitude with isospin $I$ and $q(p)$ is the relative momentum
of the intermediate (final) nucleons in the diagram $M^{}_R$.
To simplify the calculations, we take into
account only the $s$-wave $KN$ scattering in the amplitude $M^{}_R$.
We shall comment this approximation below in Sect.~4. In this case
the operator $\hmm^{}_{K\!N}$ contains no spin matrices $\bfsig$ and
Tr$\{\hPsi\bftau\hmm^{}_{K\!N}\}=0$, since the spin trace
Tr$\{\hPsi\}\sim {\rm Tr}\{\sigma\}=0$. It means that the term,
proportional to $f^{(1)}_{N\!N}$, vanishes in the amplitude $M^{}_R$.
Thus, in our approximation the NN~FSI takes place only in the triplet
$NN(^3 S_1)$ state with isospin $I=0$ in the reaction $\kdpn$ and is
absent in the reaction $\kdpp$. For the first term of the right-hand
part of Eq.~(\ref{101}) we get
\be
{\rm Tr}\{\hPsi^{\,}\bfsig\hmm^{}_{K\!N}\}\,
(\varphi^+_1\bfsig\varphi^c_2)=8\pi\sqrt{s}^{}_{K\!N}
\left[3f^{(1)}_0+f^{(0)}_0\right] (\varphi^+_1\hPsi^{\,}\varphi^c_2)
\label{102}\ee
(here an additional factor 2 comes from isospin trace Tr$\{I\}=2$),
where $f^{(I)}_0$ are the $s$-wave $KN$ amplitudes with isospins
$I\!=\!0,1$ (see Eqs.~(\ref{3})).

The "off-shell" $NN$ amplitude $f^{(0)\,off}_{N\!N}(q,p)$ is choosen
here in the form, corresponding to the scattering on the separable
Yamaguchi potential. For the "on-shell" amplitude $f^{(0)}_{N\!N}(p)$
we use the known parameters~\cite{Landau}. Then
\be
f^{(0)\,off}_{N\!N}(q,p)=\frac{p^2+\beta^2}{q^2+\beta^2}\,
f^{(0)}_{N\!N}(p),~~~~
f^{(0)}_{N\!N}(p)=\frac{1}{1/a^{(0)}_{N\!N}+\half r^{}_0 p^2-ip},
\label{11}\ee
$$
\beta\approx 240~MeV,~~~ a^{(0)}_{N\!N}=-5.4\,fm,~~~ r^{}_0=1.7\,fm.
$$
Finally, applying Eqs.~(\ref{10}), (\ref{101}), (\ref{102}) and
(\ref{11}), we obtain the amplitude $M^{}_R$ in the form
$$
M_R=c^{\,}A^{}_R\,\varphi^+_1
\left[\hat L(-p^2)-\hat L(\beta^2)\right]\varphi^c_{2\,},~~~
A^{}_R=\frac{1}{4} f^{(0)}_{N\!N}(p)_{\,}
\left[3f^{(1)}_0\!+\!f^{(0)}_0\right],
$$
\be
\hat L(x)=8\pi\intdq\,\frac{\hPsi(\bfq+\bfdeL)}
{(q^2+\!x-\!i0)},
~~~\bfdeL=\frac{\bfp_1\!+\!\bfp_2}{2},
~~~\bfp=\frac{\bfp_1\!-\!\bfp_2}{2},
\label{12}\ee
where $c=\!16\pi\sqrt{m^{\,} s}^{}_{K\!N}$. Here we evaluate the $KN$
amplitudes $f^{(I)}_0$ for the target nucleon at rest in the deuteron
rest frame. The integral $\hat L(x)$ is calculated in Subsect.~4 of
Appendix, where the analitical expression is given in the case of the
Bonn~\cite{Bonn} or Paris~\cite{Paris,Paris1} DWF.

\subsection{\bf Elastic (coherent) \bfKpD\ scattering}

Close to threshold contribution of the elastic process $\kdkd$
dominates in the total \crs\ $\sigma^t_{Kd}$. We shall use a
single-scattering approximation for the elastic scattering amplitude
$M^{el}_{K^+d}$ (see the diagram in Fig.~3), neglecting relatively
small contributions of kaon rescattering and meson exchange
currents~\cite{MEC} to the integrated \crs\ $\sigma^{el}_{K^+d}$.
In our notations the amplitude $M^{el}_{K^+d}$ reads
\be
M^{el}_{K^+d}=2\!\intdp\,{\rm Tr}\{\hPsi^+_2(\bfq_2)\hmm^{}_{K\!N}
\hPsi^{}_1(\bfq_1)\}~~ (\bfq^{}_1\!=\!\bfp-\frac{\bfpp}{2},~~
\bfq^{}_2\!=\!\bfp-\frac{\bfpp^{}_1}{2}),
\label{13}\ee
where $\hPsi^{1,2}(\bfq^{}_{1,2})$ are the operators in the
DWF~(\ref{5}) of the initial and final deuteron, respectively;
$\hmm^{}_{K^+\!N}=\!8\pi\sqrt{s}^{}_{K\!N} \hf^{}_{K^+\!N}$  is the
$KN$-scattering operator (the particle momenta are denoted in Fig.~3).

The expression for the $\kdkd$ differential \crs\ $\dso$ can be found,
for example, in Refs.~\cite{Ste64,Gia74} for the $s$-wave DWF, and in
Ref.~\cite{Gla77} for the case with the $d$-wave part of DWF included.
For the unpolarized particles this \crs\ is
$\dso\!=\!\overline {|M^{el}_{K^+d}|^2}/(8\pi\sqrt{s}^{}_{K\!d})^2$,
where $\Omega$ is the CM solid angle, and can be written as
\be
\frac{d\sigma}{d\Omega}=\frac{4s^{}_{KN}}{s^{}_{Kd}}\,
\left[|A_p\!+\!A_n|^2\left[\Phi^2_S(q)+\Phi^2_Q(q)\right]
+\frac{2}{3}|B_p\!+\!B_n|^{2\,}\Phi^2_M(q)\right].
\label{14}\ee
Here: $\bfq=\half(\bfk\!-\!\bfk^{}_1)$; $\sqrt{s}^{}_{K\!d}$ is the
total CM energy; $A^{}_{p,n}$ and $B^{}_{p,n}$ are the coefficients in
the $K^+N$ amplitudes
$\hf^{}_{p,n}\!=\!A^{}_{p,n}\!+\!B^{}_{p,n}(\bfn\bfsig)$ (we calculate
them at fixed momentum $\bfp\!=\!\half(\bfpp^{}_1\!+\!\bfpp^{}_2)$ of
the intermediate nucleon Fig.~3). The form factors $\Phi^{}_S(q)$,
$\Phi^{}_Q(q)$ and $\Phi^{}_M(q)$ in Eq.~(\ref{14}) are
\be
\Phi^{}_S=F_a+F_b,~~~~ \Phi^{}_Q=2F_c-\frac{F_d}{\sqrt{2}},~~~~
\Phi^{}_M=F_a-\frac{F_b}{2}+\frac{F_c}{\sqrt{2}}+\frac{F_d}{2},
\label{15}\ee
where
\be
\begin{array}{ll}
\dist F_a=4\pi\!\int\!dr\,j^{}_0(qr)u^2(r),~~ &
\dist ~~F_b=4\pi\!\int\!dr\,j^{}_0(qr)w^2(r), \\
\rpttt\dist F_c=4\pi\!\int\!dr\,j^{}_2(qr)u(r)w(r),~~ &
\dist ~~F_d=4\pi\!\int\!dr\,j^{}_2(qr)w^2(r).
\end{array}
\label{16}\ee
Here: $u(r)$ and $w(r)$ are the $s$- and $d$-wave components of the DWF
in coordinate representation, normalized as
$4\pi\!\int\!dr\left[u^2(r)+\!w^2(r)\right]=\!1$; $j^{}_0$ and $j^{}_2$
are the zeroth and second order spherical Bessel functions,
respectively. In the case of the Bonn~\cite{Bonn} or
Paris~\cite{Paris,Paris1} DWF the analitical expressions for the
integrals~(\ref{16}) are given in Subsect.~5 of Appendix.

\section{\bf Cross sectons of the \bfKD\ reactions}

\subsection{\bf The results of calculations and comparison
                with the data}

Here we present the results of our calculations of the $K^+d$ \crss,
based on the formulas from Sects.~2 and 3. We use the $KN$ phase shifts
from Ref.~\cite{Gibbs1} and take into account only $s$- and $p$-wave
$KN$ amplitudes, while $d$- and $f$-wave contributions are negligibly
small at the energies of interest and are neglected. We also use the
DWF of the Bonn potential~\cite{Bonn} (full model) with $s$- and
$d$-wave components included.

Fig.~4 shows the total \crss\ of the reactions $\kdpn$~($a$) and
$\kdpp$~($b$). The symbols corresponds to the experimental data from
Refs.~\cite{Gia72} (Fig.~4$a$) and \cite{Gia72,Dam75,Sak76,SLa61}
(Fig.~4$b$). The curves show the results of calculations.
In Fig.~4$a$ the curves show the contributions of the amplitudes
$M^{}_1\!+\!M^{}_2$ (dashed), $M^{}_R$ (dotted),
$M^{}_1\!+\!M^{}_2\!+\!M^{}_R$ (solid),
$M^{}_1\!+\!M^{}_2\!+\!M^{}_R$ with the $s$-wave DWF (dashed-dotted),
$M^{}_1\!+\!M^{}_2\!+\!M^{}_R$ with the $s$-wave $KN$ amplitudes (dashed
curve "S"). In Fig.~4$b$ the curves show the contributions of the
amplitude $M^{}_1\!+\!M^{}_2$. Here are also given the results,
obtained with $s$-wave DWF (dashed-dotted curve) and with the $s$-wave
$KN$ amplitude (dashed curve "S").  Comparing the solid and
dashed-dotted curves in Fig.~4, one finds that the influence of the
deuteron $d$ wave on the results is very small.

Fig.~4$a$ shows that the contribution of the NN~FSI amplitude $M^{}_R$
essentially affects the calculated \crs. The term $M^{}_R$
destructively interferes with the pole amplitudes $M^{}_{1,2}\,$ and
decreases the \crs\ at $\PLA < 200$~MeV$/c\,$ by several times.
Remember that the diagram $M^{}_R$, in which we take into account only
the $s$-wave $KN$ scattering, contains the $NN$ rescattering only in
the triplet $^3S^{}_1$ state ($S^{}_{N\!N}\!=\!1$, $I\!=\!0$). However,
the $NN$-scattering length in the singlet $^1S^{}_0$-state
($S^{}_{N\!N}\!=\!0$, $I\!=\!1$) $a^{(1)}_{N\!N}\!=24~fm$~\cite{Landau}
and is large in comparison with the triplet value $a^{(0)}_{N\!N}$ (see
Eqs.~(\ref{11})). Thus, one needs the arguments to neglect the
$NN$($^1S^{}_0$) rescattering.

This approximation should be reasonable in the momentum range where the
$p$-wave $KN$ amplitude is small. Comparing the solid curve and the
dashed one, marked by "$S$", in Fig.~4$a$, we find the influence of the
$p$-wave $KN$-scattering in the reaction $\kdpn$ to be small at
$\PLA < 250$~MeV$/c$. From this indirect estimation we expect that the
$NN$($^1S^{}_0$)-rescattering correction is not essential in this range,
but this approximation is less reliable at larger momenta $\PLA$.
In more accurate calculation of the NN~FSI correction the $p$-wave
$KN$ amplitude and $NN$($^1S^{}_0$) interaction should be also included.
We postpone this to a future study.

\vspace{1mm}
Figs.~5$a$ and 5$b$ show the total ($\stot$) and the elastic ($\sela$)
$K^+d$ \crss, respectively. The symbols are the experimental data from
Refs.~\cite{Bow70,Bow73,Bug68,Gia72,Dam75} (Fig.~5$a$) and
\cite{Gia72,Sak75} (Fig.~5$b$). Here, the calculated \crs\ $\stot$ is
taken as a sum of the $\kdpn$, $\kdpp$ and $\kdkd$ \crss, shown by the
solid curves in Figs.~4$a$, 4$b$ and 5$b$. The dashed-dotted curves in
Fig.~5 are the results, obtained with the $s$-wave DWF. The dashed curve
in Fig.~5$a$ shows the result for $\stot$ in which the contribution of
the $\kdpn$ \crs\ is not corrected for NN~FSI.

Let us comment here the results of Ref.~\cite{Gibbs}, where data
on the total \crs\ $\stot$ were analysed and $\stot$ was evaluated
through the unitarity with single- and double-scattering $\kdkd$
amplitudes employed. The unitarity cuts of the single- and
double-scattering terms correspond to the contributions (summed over
the $K^+pn$ and $K^0pp$ channels) of the squares
$|M^{}_1|^2\!+\!|M^{}_2|^2$ and of the interference term
2Re$(M^*_1 M^{}_2)$, where $M^{}_{1,2}$ are the pole amplitudes.
Our comment is the following.

1.~The \crs\ $\stot$, calculated in Ref.~\cite{Gibbs}, does not include
the contribution of the elastic $K^+\!d$ scattering, which dominates at
low momenta $\PLA < 100$~MeV$/c$. Results of our computations at
$\PLA = 100$~MeV$/c$ are shown in Figs.~4 and 5. They give
$\sela =\!34.4~mb$, $\sine =\sigma(K^+pn+K^0pp)=\!0.9~mb$ $(9.7~mb)$ and
$\stot =35.3~mb$ $(44.1~mb)$ with (without) NN~FSI diagram $M^{}_R$
included. On the other hand, in Ref.~\cite{Gibbs} (see Fig.~5 there) one
finds $\stot =\sine\approx 27~mb$. This value of $\sine$ is too large
due to the following reasons. Firstly, the elementary $KN$ amplitude,
used in Ref.~\cite{Gibbs}, corresponds to the real (not virtual) target
nucleon (see Eq.~(17) there). Secondly, they average the $K^+N$ \crs\
over the Fermi momentum distribution (see Eq.~(19) there) in the range
$0\!<\!p\!<\!\infty$, neglecting the kinematical boundaries. Thus, the
\crs\ in Ref.~\cite{Gibbs} is overestimated in the nearthreshold region,
where the kinemaical boundaries are important.

2.~The double-scattering $K^+d$ amplitude is considered in
Ref.~\cite{Gibbs} under the following assumptions. The propagator of
the intermediate kaon is taken in a "static nucleon" approximation.
Thus, the contribution of this diagram to $\stot$ depends on the energy
like 2-particle phase space instead of 3-particle ($K\!N\!N$) one as it
should be. The elementary $K^+N$ amplitudes modifyed by the
$K^+N(I\!=\!0)$-resonance contribution are taken out of the integral
over the momenta in the intermediate $K\!N\!N$ state and taken at fixed
nucleon momenta. This approximation is widely used for the hadronic
amplitudes, usually being smooth functions in comparison with the rapid
$p$ dependence of the nuclear wave functions. However, in the case of a
narrow ($\Gamma\sim 1$~MeV) $K^+N$ resonance this approximation can be
not reliable.

Summarising the results of this Sect., we conclude that the approach,
based on the pole diagrams and modifyed by the NN~FSI term (simplifyed
as discussed above), gives a reasonable description of the existing
data on the integrated $K^+d$ \crss\ in the range $\PLA < 800$~MeV$/c$.
The NN~FSI effect is found to be large. It would be useful to have the
data on the $\stot$ and $\kdpn$ \crs\ in the range, say
$\PLA < 400$~MeV$/c$ (where they are absent now), for more detailed
study of the NN~FSI effect and comparison with the data.

\subsection{\bf On the extraction of the isoscalar $\bfKNI$
             scattering length}

To determine the isoscalar $KN$ scattering length $a^{}_0$ one needs
the additional data on the kaon-neutron scattering, but the neutron
targets do not exist. Thus, to extract the value of $a^{}_0$, one
should compare the theoretical predictions with the data on the \crss\
for the existing targets and the deuteron one is preferable.
As a source of slow kaons the decay $\phi(1020)\!\to\!K^+K^-$ at
rest can be used and the $\phi(1020)$ mesons can be produced in the
$e^+e^-$-collisions at DA$\Phi$NE accelerator in Frascatti.

Fig.~6 show our predictions for the $K^+d$ \crss\ at the initial
momentum $\PLA =127$~MeV$/c$, which is the kaon momentum in the
$\phi(1020)$ decay. At this momentum the $p$-wave $KN$ amplitudes
are negligibly small and we use here only the $s$-wave amplitudes.
Figs.~6$a$ and 6$b$ show the $\kdpn$ ($a$) and $\kdpp$ ($b$) \crss.
The total and the elastic $K^+d$ \crss\ are given in Figs.~6$c$ and
6$d$, respectively. The results are presented as functions of $a^{}_0$
in some range around the "realistic" values, given in the Table. The
results are given for two fixed values $a^{}_1=-0.328~fm$ (curves~1)
and $a^{}_1=-0.308~fm$ (curves~2), taken from the Table. The solid and
dashed curves in Figs.~6$a$ and 6$c$ show the results obtained with and
without NN~FSI, taken into account. Thus, the NN~FSI amplitude strongly
affects the $\kdpn$ and total $K^+d$ \crss\ for slow kaons.

\section{\bf Conclusion}

The theoretical predictions for the $K^+d$ \crss\ were presented in
the "quasi-elastic" energy range $\PLA < 0.8$~GeV$/c$, where the
particle-production processes in the elementary $KN$ interactions can
be neglected. We used the approach, which employs the pole $\KDNN$
amplitudes, the NN~FSI correction, and the "realistic" $KN$ phase
shifts. In our approximation we neglected the $p$-wave $KN$ scattering,
when calculating the NN~FSI correction. Since the $s$-wave
$KN$-scattering amplitude is non-spin-flip the NN~FSI takes place only
in the $NN(^3 S_1)$ state, forbidden for the reaction $\kdpp$ due to
the Pauli principle. This approximation should be reasonable in the
low-momentum range (estimated in Sect.~4), where the $KN$-scattering
amplitude is predominantly $s$-wave.

A reasonable description of the data on the integrated $\kdpn$ and
$\kdpp$ \crss\ as well as on the total and elastic $K^+d$ \crss\ were
obtained. The NN~FSI diagram affects strongly the value of the $\kdpn$
\crs\ in the low energy region and interferes destructively with the
pole diagrams. However, data on the $\kdpn$ \crs\ are available only at
higher energies $\PLA >600$~MeV$/c$, where our calculations of NN~FSI
are less reliable and some defects of the theoretical description are
seen. It would be interesting to measure the $\kdpn$ \crs\ at low
enegies, say $\PLA < 400$~MeV$/c$, where the NN~FSI effect is large,
to investigate the role of this mechanism.

Predictions were also given for the integrated \crss\ of the $K^+d$
reactions with slow kaons as functions of the isoscalar
$KN$-scattering length $a^{}_0$. These results would be usefull
for extraction of the $a^{}_0$ value from the data. The
corresponding experiments with slow kaon beam from the $\phi(1020)$
decays may be proposed, say, for the DA$\Phi$NE accelerator. At this
energy ($\PLA =127$~MeV$/c$) the NN~FSI effect is very strong in
the reaction $\kdpn$ as it is seen from Fig.~4$a$. Thus, study of
this reaction at the DA$\Phi$NE machine would be very important.

In the more detailed treatment of NN~FSI the $p$-wave $KN$ scattering
and $NN(^1 S_0)$ interaction should be also taken into account. We
postpone this for a future study. However, these improvements should
not affect essentially our results for the $K^+d$ \crss\ at the energy
of kaons from $\phi(1020)$ decay and the conclusion about the large
magnitude of the NN~FSI effect.

Authors acknowledge support of the Federal Agency of Atomic Energy of
Russian Federation. Participations of V.E.T. and A.E.K. were supported
by DFG-RFBR grant no. 05-02-04012 (436 RUS 113/820/0-1(R)).

\vspace{3mm}
\centerline{\bf Appendix}

\vspace{2mm}
{\bf 1.~Cross sections and phase spaces.}
\vspace{1mm}

Calculating the $Kd$ \crss, we use the invariant amplitudes
with Feynman normalization. The \crs\ $\sigma_n$
of the process $\,a\!+\!b\!\to\!1+\!\cdots\!+\!n\,$ reads
$$
\sigma_n=\frac{1}{4q^{}_{ab}\sqrt{s}}\int\! |M|^2 d\tau_n,
~~~~
d\tau_n=I_n (2\pi)^4\delta^{(4)}(P_i-P_f)
\prod^n_{i=1} \frac{d^3 \bfp_i}{(2\pi)^3 2E^{}_i}.
\eqno{(\rm A.1)}
$$
Here $M$ is the invariant amplitude; $\sqrt{s}$ is the total CM energy;
$\,q_{ab}$ is the initial relative momentum and
$q^{}_{ab}\!=\!\lambda(s,m^2_a,m^2_b)$, where $\lambda(z,x,y)\!=\!
\sqrt{(z\!-\!x\!-\!y)^2\!-\!4xy}^{\,}/2\sqrt{z}$ and $m^{}_a(m^{}_b)$
is the mass of the particle $a(b)$; $\,d\tau_n$ is the element of the
final $n$-particle phase space; $P_i(P_f)$ is the total initial (final)
four-momentum; $E^{}_i\,$ and $\bfp_i$ are the total energy and
momentum of the $i$-th final particle; the factor
$I_n\equiv 1/n_1!\cdots n_k!$ takes into account the identity of final
particles, where $n_i$ is the number of particles of the $i$-th type
($n_1\!+\!\cdots\!+\!n_k\!=\!n$). Then, the \crs\ $\sigma$ of the
reaction $\KDNN$ with unpolarized particles can be written as
$$
\sigma=\!\!\int\!\!\frac{d\sigma}{d\snn}\,d\snn,~~
\frac{d\sigma}{d\snn}=\frac{I_n}{2(4\pi)^4 Q^{\,} s}
\int\overline {|M|^2}\,Q^{}_1\,p\,dz^{}_1 dz d\varphi,
\eqno{(\rm A.2)}
$$
where $z\!=\!\cos\theta$, $z^{}_1\!=\!\cos\theta^{}_1\,$. Here:
$\overline {|M|^2}$ is the square of the reaction amplitude $M$ with
unpolarized particles; $Q\!=\!|\bfqq|$ and $Q^{}_1\!=\!|\bfqq^{}_1|$,
where $\bfqq$ and $\bfqq^{}_1$ are the CM momenta of the incoming and
outgoing kaon, respectively; $\theta^{}_1$ is the CM polar angle of the
outgoing kaon; $p\!=\!|\bfp|$; $\bfp$, $\theta$ and $\varphi$ are the
momentum and the angles (polar and azimuthal) of the outgoing nucleon,
say $N^{}_{1\,}$, in the $NN$ rest frame.

\vspace{2mm}
{\bf 2.~Kinematics.}
\vspace{1mm}

Let us express all the momenta used to calculate the $\KDNN$ amplitude,
through the variables $\snn$, $z^{}_1$, $z$ and $\varphi$ of the
integrals~(A.2). One can write
$$
Q=\lambda(s,m^2_K,m^2_d),~~~ Q^{}_1=\lambda(s,m^2_K,s^{}_{N\!N}),
~~~ p=\sqrt{m(\snn\!-\!2m)}, \eqno{(\rm A.3)}
$$
where $m^{}_K$ ($m^{}_d$) is the kaon (deuteron) mass and the function
$\lambda(\cdots)$ is defined after Eq.~(A.1). Let us introduce the
notations: $\bfp'^{}_{1,2}$ and $\bfp^{}_{1,2}$ -- the momenta of the
final nucleons in the reaction and in the deuteron rest frames,
respectively; $\bfq'^{}_{1,2}$ and $\bfq^{}_{1,2}$ -- the initial and
final nucleon momenta, respectively, in the rest frame of the
$KN^{}_{1,2}\!\to\!KN^{}_{1,2}$ subprocess in the diagram
$M^{}_{1,2}$ (Fig.~2). Then we write:
$$
\bfp'^{}_{1,2}=\pm\bfp-\frac{\bfqq^{}_1}{2},~~
\bfp^{}_{1,2}=\bfp'^{}_{1,2}\!+\!\frac{\bfqq}{2},~~
\bfq^{}_{1,2}=\frac{\omega^{\,}\bfp'^{}_{1,2}\!-\!m^{\,}\bfqq^{}_{1\,}}
{m+\omega},~~ \bfq'^{}_{1,2}=\bfq^{}_{1,2}\!+\!\bfqq^{}_1\!-\!\bfqq,
\eqno{(\rm A.4)}
$$
where $\omega\!=\!\sqrt{m^2_K\!+\!p^2_{_{\rm LAB}}}$ is the kaon total
energy. The values $q^{}_i\!=\!|\bfq^{}_i|$,
$z^{}_{KN_i}\!=\!(\bfq'^{}_i\bfq^{}_i)/q'^{}_i q^{}_i$ and $\bfn^{}_i\!
=\![\bfq'^{}_i\!\times\!\bfq^{}_i]/|[\bfq'^{}_i\!\times\!\bfq^{}_i]|$
are used to calculate the $K^+N^{}_i$-scattering amplitude ($i=1,2$),
according to Eqs.~(\ref{3}).

\vspace{2mm}
{\bf 3.~The square of the $\bfKDNN$ amplitude}
\vspace{1mm}

Here we write out the square $\overline {|M|^2}$ of the amplitude
$M\!=\!M^{}_1\!+\!M^{}_2\!+\!M^{}_R$ for unpolarized particles,
applying the formulas from Sects.~2 and 3. Hereafter, we exclude the
isospin variables and fix the nucleons with momenta $\bfp^{}_1$ and
$\bfp^{}_2$ in the reaction $\kdpn$ as proton and neutron, respectively.
Then, $I_n\!=1(\half)$ in Eq.~(A.2) for the reaction $\kdpn$ ($\kdpp$).
Let $A_{(i)}$ and $B_{(i)}$ be the coefficients in the $K^+N_i$
amplitude $\hf^{}_{K\!N_i}\!=\!A_{(i)}\!+\!B_{(i)}(\bfn^{}_i\bfsig)$.
Then: $A_{(1)}\!=\!A_1$ and $B_{(1)}\!=\!B_1$
($A_{(2)}\!=\!\half(A_1\!+\!A_0)$ and $B_{(2)}\!=\!\half(B_1\!+\!B_0)$)
for the $K^+p\,$($K^+n$)-scattering subprocess in the diagram
$M^{}_1$($M^{}_2$) for the reaction $\kdpn$;
$A_{(i)}\!=\!\half(A_1\!-\!A_0)$ and $B_{(i)}\!=\!\half(B_1\!-\!B_0)$
for the charge exchange subprocess $K^+n\!\to\!K^0 p$ in the reaction
$\kdpp$; the values $A^{}_I$ and $B^{}_I$ are given in Eqs.~(\ref{3}).
For the different terms of the expression $\overline {|M|^2}$ we obtain
(hereafter Tr$^{\,}\{(\cdots)\}$ means the trace of the
2$\times$2 matrix $(\cdots)$ with spin indices)
$$
\begin{array}{ll}
\overline {|M^{}_i|^2}\!&= c^{2\,}{\rm Tr}\{\hPsi(\bfp_j)\hPsi^+(\bfp_j)
\hf^+_{K\!N_i}\hf^{}_{K\!N_i}\} ~~(i\ne j\!=\!1,2),
\\ \rppt
\overline {M^{}_1 M^+_2}\!&=\pm\, c^{2\,}{\rm Tr}\{\hPsi(\bfp_2)
(A^{}_{(2)}\!-\!B^{}_{(2)}\hat n^{}_2)^{+\,}\hPsi^+(\bfp_1)
(A^{}_{(1)}\!+\!B^{}_{(1)}\hat n^{}_1)\},
\\ \rppt
\overline {M^+_iM^{}_R}\!&\!= c^2 A^{}_{R\,} {\rm Tr}
\{\hPsi^+(\bfp_j)\hf^+_{K\!N_i}\hat L\} ~~(i\!\ne\!j\!=\!1,2),
\\ \rppt
~~\overline {|M^{}_R|^2}\!&\!=c^{2\,}|A^{}_R|^{2\,}{\rm Tr}\{\hat L
\hat L^+\},~~ \hat L\!=\!\hat L(-p^2)\!-\!\hat L(\beta^2),
\end{array}
\eqno{(\rm A.5)}
$$
where $\hat n_i=(\bfn_i\bfsig)$. Here: the sign "$-$" in the expression
for $\overline {M^{}_1 M^+_2}$ corresponds to the case of the reaction
$\kdpp$; the terms $\overline {M^+_iM^{}_R}$ and
$\overline {|M^{}_R|^2}$ are given for the reaction $\kdpn$; the
quantities $A^{}_R$ and $\hat L(x)$ are given by Eqs.~(\ref{12});
calculating the factors $c^{}_i$ and $c$, defined in Eqs.~(\ref{7}) and
(\ref{12}), we use the value of $s^{}_{K\!N}$ for the nucleon at rest,
i.e. $c^{}_i\!=\!c=16\pi\sqrt{m^{\,} s}^{}_{K\!N}$ and
$s^{}_{K\!N}\!=\!m^2_K\!+\!m^2\!+\!2\omega m$.

Let us introduce the functions $f(p)$ and $g(p)$, rewriting
$\hPsi(p)$~(\ref{5}) in the form
$$
\begin{array}{c}
{\dist \hPsi(\bfp)\!=\!f(p)(\bfeps\bfsig)+g(p)
\frac{(\bfp\bfeps)(\bfeps\bfsig)}{p^2}},
\\ \rpttt
{\dist f(p)\!=\!\frac{u(p)}{\sqrt{2}}+\frac{w(p)}{2},~~
g(p)\!=\!-\frac{3w(p)}{2}}.
\end{array}
\eqno{(\rm A.6)}
$$
Finally, from Eqs.~(A.5) and (A.6), we obtain
$$
\begin{array}{ll}
\overline {|M^{}_i|^2}\!&= c^2\,[u^2(p^{}_j)\!+\!w^2(p^{}_j)]\,
[^{\,}|A^{}_{(i)}|^2 + |B^{}_{(i)}|^2] ~~~(i\ne j\!=\!1,2),
\\ \rpttt
{\rm Re}\,\overline {(M^{}_1 M^+_2)}\!&=\pm\, c^2\biggl[
{\rm Re}^{\,}(A^{}_{(1)}A^*_{(2)})^{\,}\biggl(u(p_2)u(p_1)\!+\!
w(p_2)w(p_1) {\dist \frac{3z^2_p\!-\!1}{2}}\biggr)
\\ \rpttt
 &+{\dist \frac{2g_{1\,} g_{2\,} z_p}{3p_{1\,} p_2}}\biggl(
(\bfp_2[\bfn_2\!\times\!\bfp_1])^{\,}{\rm Im}^{\,}(A^{}_{(1)}B^*_{(2)})
\!-\!(\bfp_1[\bfn_1\!\times\!\bfp_2])^{\,}{\rm Im}^{\,}
(B^{}_{(1)}A^*_{(2)})\biggr)
\\ \rpttt
 &+{\rm Re}^{\,}(B^{}_{(1)}B^*_{(2)})^{\,} \biggl[{\dist \frac{2}{3}}
(\bfn_1\bfn_2)\Bigl(f_1f_2\!+\!g_1f_2\!+\!f_1g_2\!+\!g_1g_{2\,}z^2_{p\,}
\Bigr)
\\ \rpttt & {\dist +\frac{4(\bfp_1\bfn_1)(\bfp_2\bfn_2)}{3}
\biggl(\frac{g_1f_2}{p^2_1}
+\frac{f_1g_2}{p^2_2}-\frac{g_{1\,}g_{2\,}z_p}{p_{1\,} p_2}\biggr)
\biggr]\biggr]~~~ \Bigl( z_p=\frac{(\bfp_1\bfp_2)}{p_1 p_2}\Bigr)},
\end{array}
$$
$$ \eqno{(\rm A.7)} $$
$$
\begin{array}{ll}
{\rm Re}\,\overline {(M^+_iM^{}_R)}\!&\!= c^{2\,}\Biggl[
{\dist 2^{\,}{\rm Re}^{\,} \biggl(A^{}_{R\,}A^*_{(i)}\biggl[
\Bigl(f_j+\frac{g_j}{3}\Bigr) A +\frac{g_j}{3} B^{\,}\biggl(
\frac{3(\bfp_j\bfdeL)^2}{p^2_{j\,}\Delta^2}-1\biggr)\biggr]\biggr)}
\\ \rpttt
 &\!+ {\dist {\rm Im}^{\,}(A^{}_{R\,} B^*_{(i)\,}B)\, g_j
\frac{(\bfp_j\bfdeL)}{p^2_{j\,}\Delta^2}(\bfp_i[\bfn_i\!\times\!\bfp_j])
\Biggr] }~~ (i\!\ne\!j\!=\!1,2),
\\ \rpttt
\overline {|M^{}_R|^2}\!&=2c^{2\,}|A^{}_R|^{2\,}(|A|^2+2|B|^2),
\end{array}
$$
where $f_{1,2}\!=\!f(p_{1,2})$, $g_{1,2}\!=\!g(p_{1,2})$. Here
$A$ and $B$ are the coefficients in the expression for $\hat L(x)$,
given below in Subsect.~4. If one neglects the $d$-wave part of DWF,
Eqs.~(A.7) give
$$
\begin{array}{ll}
\overline {|M^{}_i|^2}\!&= c^2\,u^2(p^{}_j)\,
[^{\,}|A^{}_{(i)}|^2 + |B^{}_{(i)}|^2] ~~~(i\ne j\!=\!1,2),
\\ \rppt
\overline {M^{}_1 M^+_2}\!&=\pm\, c^{2\,} u(p^{}_1) u(p^{}_2)\,
\Bigl[ A^{}_{(1)} A^*_{(2)}\!+\!\frac{1}{3} B^{}_{(1)} B^*_{(2)}
(\bfn^{}_1\bfn^{}_2)\Bigr],
\\ \rppt
\overline {M^+_iM^{}_R}\!&= c^{2\,}
A^{}_{R\,}A^*_{(i)}A^{\,}u(p^{}_j),~~~
 \overline {|M^{}_R|^2}\!=2c^{2\,}|A^{}_R|^{2\,}|A|^2.
\end{array}
\eqno{(\rm A.8)}
$$

\vspace{2mm}
{\bf 4.~Calculation of the operator\ {\mbox {\boldmath $\hat L(x)$}}.}
\vspace{1mm}

Here, it is convenient to use the DWF in coordinate representation, i.e.
$$
\hPhi(\bfr)=\!\frac{u(r)}{r\sqrt{2}}(\bfeps\bfsig)-\!\frac{w(r)}{2r}
\left[\frac{3(\bfr\bfeps)(\bfr\bfsig)}{r^2}-\!(\bfeps\bfsig)\right],
~~\hPsi(\bfq)=\int\!d^3r\,e^{-i\bfq\bfr}\hPhi(\bfr),
\eqno{(\rm A.9)}
$$
where $\hPsi(\bfq)$ is given by Eqs.~(\ref{5}). For the DWF's of
Bonn~\cite{Bonn} and Paris~\cite{Paris} potentials the $s$- and
$d$-wave functions ($u$ and $w$, respectively) were
parametrized \cite{Bonn,Paris1} in the form
$$
u(p)=\!\sum_i \frac{C_i}{p^2\!+\!m^2_i},~~
w(p)=\!\sum_i \frac{D_i}{p^2\!+\!m^2_i}~~
(\sum_i C_i=\!\sum_i D_i=\!\sum_i D_i m^2_i
=\!\sum_i \frac{D_i}{m^2_i}=\!0),
$$ $$
u(r)=\!\sum_i \frac{C_i}{4\pi}\,e^{-m_ir},~~
w(r)=\!\sum_i \frac{D_i}{4\pi}\,e^{-m_ir}\,
\left(1+\frac{3}{m_i r}+\frac{3}{m^2_i r^2}\right),
\eqno{(\rm A.10)}
$$

\vspace{1mm}
Calculating the integral $\hat L(x)$~(\ref{12}), we transform the
factors $\hPsi(\bfq\!+\!\bfdeL)$ and $(q^2\!+\!x\!-\!0)^{-1}$ into the
$\bfr$-representation and obtain
$$
\hat L(x)\!=\!2\int\!\frac{d^3r}{r}{\dist
e^{{\dist i\bfdeL\bfr\!+\!\alpha r}}} \hPhi(\bfr),~~~ \alpha\!=
\biggl\{\!\begin{array}{ll} -a & (x>0) \\ ~ia & (x<0) \end{array},~~~
a\!=\!\sqrt{|x|}.
\eqno{(\rm A.11)}
$$
Let us rewrite the operator $\hPhi(\bfr)$ in Eqs.~(A.9) as
$\hPhi(\bfr)\!=\!\Phi^{}_{ij}\varepsilon^{}_i\sigma^{}_j$, where
$\varepsilon^{}_i$ is the $i$-th component of the deuteron polarization
vector. Then, from Eq.~(A.11), we arrive at the expressions
$$
\hat L(x)=L^{}_{ij}\varepsilon^{}_i\sigma^{}_j,~~~~~
L^{}_{ij}=A\dij+B\biggl( {\dist
\frac{3\Delta^{}_i\Delta^{}_j}{\Delta^2}-\dij }\biggr),~~~
\eqno{(\rm A.12)}
$$ $$
\begin{array}{ll}
A\!&={\dist \sqrt{2}\int\!\frac{d^3r}{r^2}\,
e^{{\dist i\bfdeL\bfr\!+\!\alpha r}} u(r)=4\sqrt{2}^{\,}\pi
\int\!dr\, e^{{\dist \alpha r}} u(r) j^{}_0(r\Delta)},
\\ \rpttt
B\!&={\dist -\frac{1}{2}\int\!\frac{d^3r}{r^2}\,
e^{{\dist i\bfdeL\bfr\!+\!\alpha r}} w(r)\biggl(
\frac{3(\bfdeL\bfr)^2}{r^2\Delta^2}-1\biggr)=4\pi\int\!dr\,
 e^{{\dist \alpha r}} w(r) j^{}_2(r\Delta)}.
\end{array}
$$
Making use of the functions $u(r)$ and $w(r)$, given by Eqs.~(A.10),
we obtain
$$
A={\dist \sum_i\sqrt{2}C^{}_i J(m^{}_i,a,\Delta),~~~ a=\sqrt{|x|}},
\eqno{(\rm A.13)}
$$ $$
B={\dist \sum_i D^{}_i\biggl[
\frac{3(\Delta^2\!+\!x\!-\!m^2_i)}{8m^{}_i\Delta^2}+
\frac{3(\Delta^2\!+\!x\!-\!m^2_i)^2\!+\!4m^2_i\Delta^2}{8m^2_i\Delta^2}
J(m^{}_i,a,\Delta)\biggr] },
$$
where
$$
\begin{array}{ll}
J(m,a,\Delta)\!&={\dist ~\frac{1}{\Delta}\,
{\rm arctan}\frac{\Delta}{m\!+\!a}~~~~~  (x>0)},
\\ \rpttt
J(m,a,\Delta)\!&={\dist \frac{1}{2\Delta}\,\biggl[{\rm arctan}
\frac{a\!+\!\Delta}{m}\!-\!{\rm arctan}\frac{a\!-\!\Delta}{m}\!+\!
\frac{i}{2}\ln\frac{m^2\!+\!(a\!+\!\Delta)^2}{m^2\!+\!(a\!-\!\Delta)^2}
\biggr]~~ (x<0)}.
\end{array}
$$

\vspace{2mm}
{\bf 5.~Expressions for the integrals
        {\mbox {\boldmath $F^{}_{a,b,c,d}$}}~(\ref{16}).}
\vspace{1mm}

For the wave functions $u(r)$ and $w(r)$, given by Eqs.~(A.10), one
can calculate the integrals $F^{}_{a,b,c,d}$~(\ref{16}) analytically,
and we obtain
$$
\begin{array}{l}
\dist F_a\!=\!\sum_{ij}\frac{C_iC_j}{4\pi\Delta}\,A^{}_{ij\,}~~~
\left(A^{}_{ij}\equiv {\rm arctan}
\frac{\Delta}{m^{}_i\!+\!m^{}_j}\right),\\
\rpttt \dist F_b\!=\!\sum_{ij}\frac{D_iD_j}{8\pi}\,
\frac{3(x\!+\!y\!+\!\Delta^2)^2-\!4xy}{4xy\Delta}\,A^{}_{ij}
~~(x=m^2_i,~~y=m^2_j), \\
\rpttt \dist F_c\!=-\!\sum_{ij}\frac{C_iD_j}{8\pi}\,
\left[\frac{3x}{4m_j\Delta^2}
+\frac{4y\Delta^2\!+\!3(x\!-\!y\!+\!\Delta^2)^2}{4y\Delta^3}
\right] A^{}_{ij\,}, \\
\rpttt \dist F_d\!=\!\sum_{ij}\frac{D_iD_j}{8\pi}\,
\frac{3(x\!+\!y\!+\!\Delta^2)[\Delta^4\!-\!(x\!-\!y)^2]\!
-\!8xy\Delta^2}{8xy\Delta^3}\, A^{}_{ij\,}.
\end{array}
\eqno{(\rm A.14)}
$$

\vspace{2mm}
{\bf 6.~Estimation of the Coulomb \crs\
        {\mbox {\boldmath $\sigma^{}_C$}}}
\vspace{1mm}

Here we estimate the pure Coulomb \crs\ of the reaction $\kdpn$ in
nonrelativistic case. With the $s$-wave DWF $u(p)$, we can write
$$
\sigma^{}_C=\intdp\,u^2(p) \int\!d\Omega\,|f^{}_c|^2,~~~~
f^{}_c=\frac{2\alpha^{}_{c\,}\mu}{t}.
\eqno{(\rm A.15)}
$$
Where $\bfp$ is the neutron-spectator 3-momentum in the deuteron rest
frame; $f^{}_c$ is the Born amplitude of the Coulomb $K^+\!p$
scattering; $d\Omega=dz d\varphi$ is the solid angle element in the
final $K^+\!p$ system; $\alpha^{}_c\approx 1/137$;
$\mu=m^{}_K m/(m\!+\!m^{}_K)$ is the reduced mass; and $t$ is the
four-momentum transfer squared. In the nonrelativistic form
$t=-(\bfq^{}_1\!-\!\bfq)^2$, where $\bfq^{}_1$ ($\bfq$) is the initial
(final) relative 3-momentum in the subprocess $K^+p^{}_1\!\to\!K^+p$
and $p^{}_1$ is the virtual proton with the mass $m^{}_1\!\ne\!m$.

For the angular part of the integral $\int\!d\Omega\,|f^{}_c|^2$,
making use of the relations
$$
q^2_1\!-\!q^2=2\mu^{\,}(m\!-\!m^{}_1),~~~~ m^{}_1
=m-\frac{p^2\!+\!\alpha^2}{m},~~~ \alpha^2\!=\!m\varepsilon^{}_d,
$$
where $\varepsilon^{}_d$ is the deuteron binding energy, we obtain
$$
\int\!d\Omega\,|f^{}_c|^2=
\frac{4\pi\alpha^2_{c\,}m^2}{(p^2+\alpha^2)^2}
\eqno{(\rm A.16)}
$$
We shall estimate the \crs\ $\sigma^{}_{C~}$(A.15) with the DWF of the
simplest form $u(p)=\sqrt{8\pi\alpha}/(p^2\!+\!\alpha^2)$.
Formally, the integral
$\int\!d^{\,3}\bfp$~(A.15) depends on the kinematical boundaries through
the condition $\sqrt{s}^{}_{K^+p} > m+m^{}_K$, but we shall calculate it
in the range $0\!<\!p\!<\!\infty$. In this approximation it is supposed
that the DWF is a rapid function of $p$ and the process $\kdpn$ is
considered in the region not very close to the threshold. However, we
take into account the $p$-dependent factor~(A.16). Finally, we obtain
$$
\sigma^{}_C=16\alpha^2_{c\,}\alpha^{\,} m^2 \int\limits^{\infty}_0
\frac{p^2 dp}{(p^2+\alpha^2)^4}~=~
\frac{\pi^{\,}\alpha^2_c}{2^{\,}\varepsilon^2_d}~\approx ~6.5~{\rm mb}.
\eqno{(\rm A.17)}
$$
This value is not very small in comparison with the hadronic $\kdpn$
\crs, shown in Fig.~4$a$. We neglect the Coulomb contribution,
since it is concentrated in the region of small scattering angles, which
may be not accepted by the detectors.


\newpage
\centerline{\bf Figure captions}
\begin{itemize}
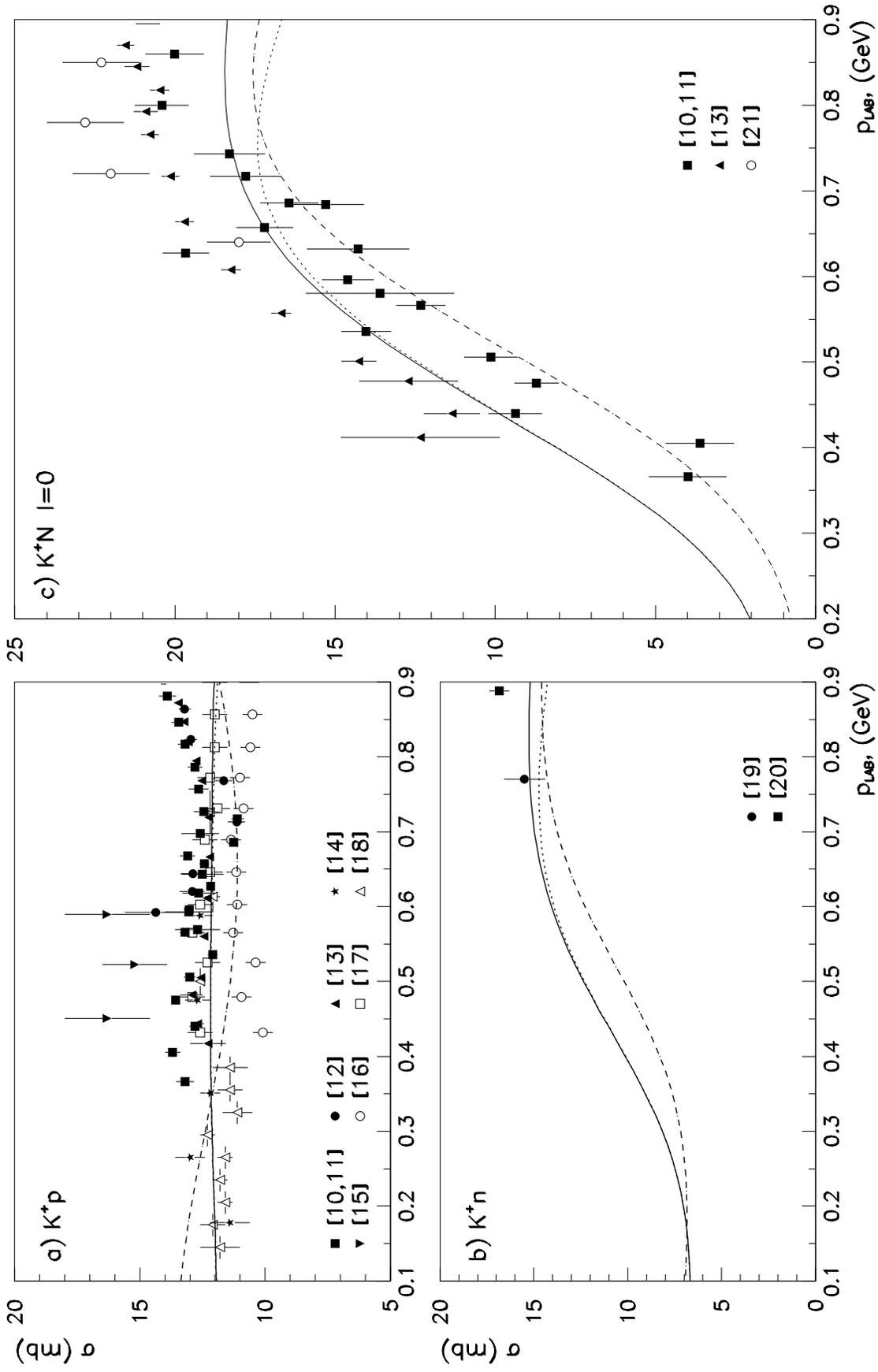
\item[Fig.~1:~]
The total $K^+p$ ($a$), $K^+n$ ($b$) and $K^+N(I=0)$ ($c$) \crss.
The curves correspond to different sets of the $KN$ phase shifts (see
text), and the symbols -- to the experimental data. The filled (empty)
symbols in Fig.~1$a$ show the data on the total (elastic) $K^+p$ \crss.

\vspace{2mm}
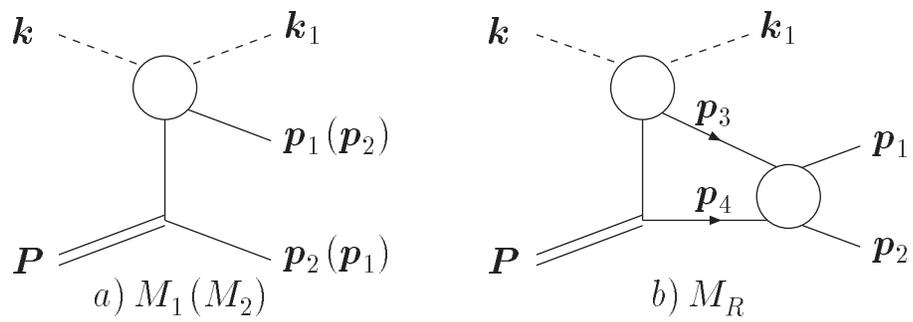
\item[Fig.~2:~]
The pole ($M^{}_1$ and $M^{}_2$) and the NN~FSI ($M^{}_R$) diagrams for
the $\KDNN$ process. The solid, dashed and double lines correspond to
the kaons, nucleons and deutrons, respectively.

\vspace{2mm}
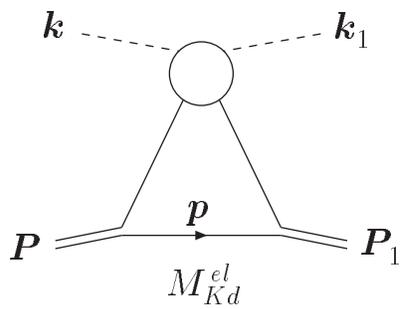
\item[Fig.~3:~]
A single-scattering diagram for the elastic process $\kdkd$.
The lines of the particles are the same as in Fig.~2.

\vspace{2mm}
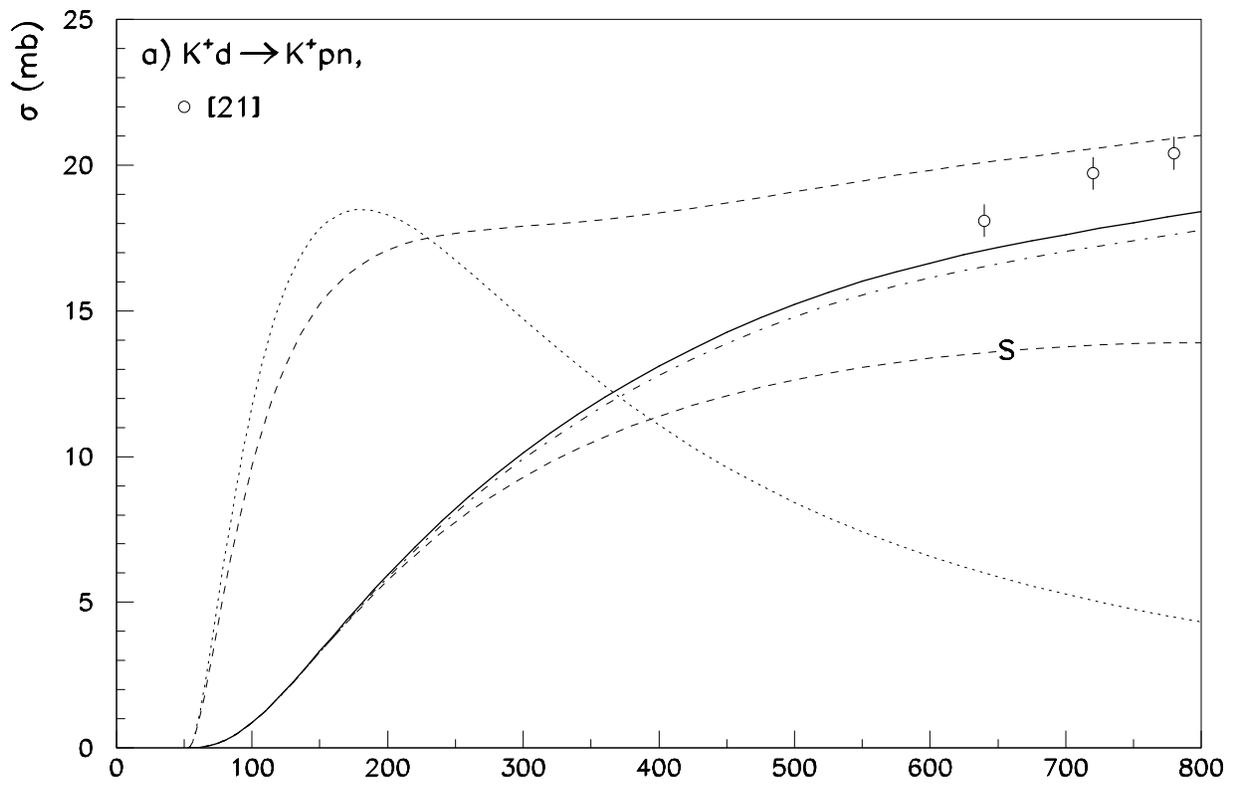
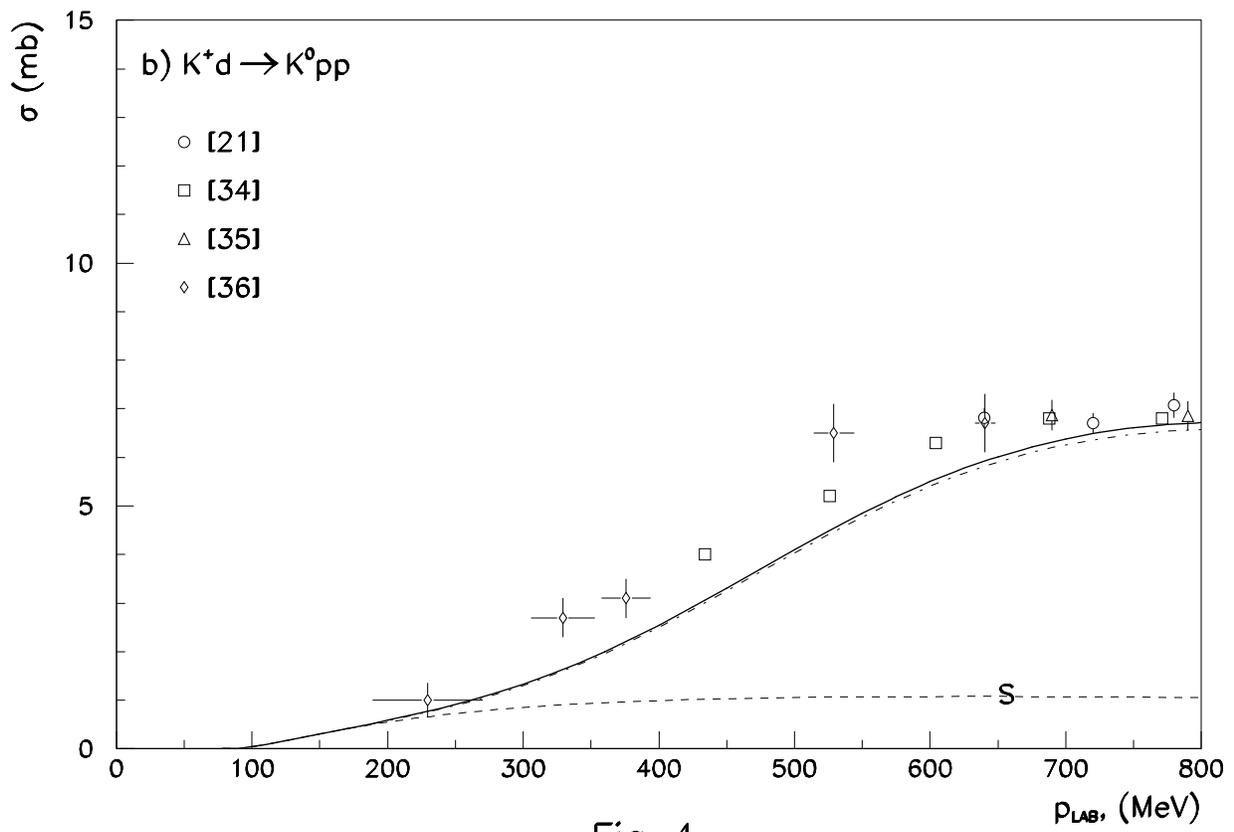
\item[Fig.~4:~]
The total \crss\ of the reactons $\kdpn$~($a$) and $\kdpp$~($b$).
The symbols correspond to the experimental data (see references on the
plot). The curves present the results of computations. Solid curves
show the contributions of the amplitudes $M^{}_1\!+\!M^{}_2\!+\!M^{}_R$
(Fig.~$a$) and $M^{}_1\!+\!M^{}_2$ (Fig.~$b$); dashed-dotted curves --
the results, obtained with the $s$-wave part of DWF; dashed curves "$S$"
-- the results, obtained with the $s$-wave $KN$ amplitudes. The dashed
and dotted curves in Fig.~4$a$ show the contributions of the amplitudes
$M^{}_1\!+\!M^{}_2$ and $M^{}_R$, respectively.

\vspace{2mm}
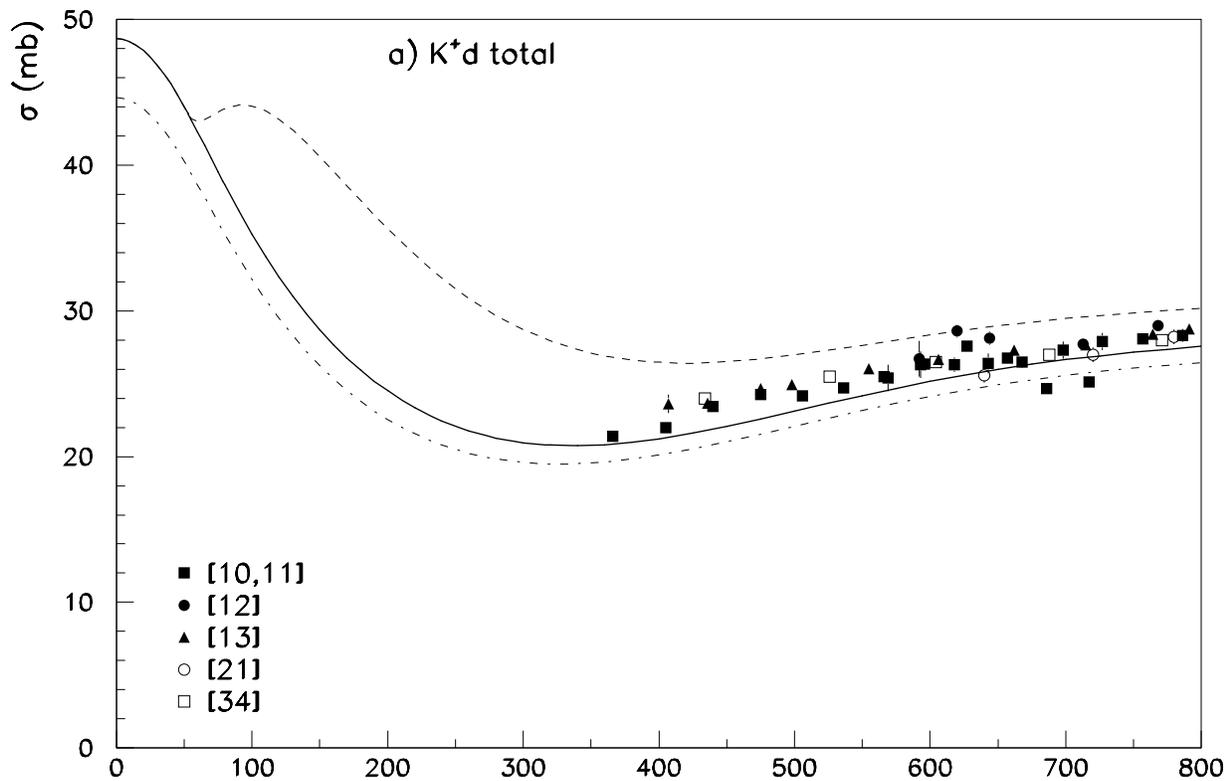
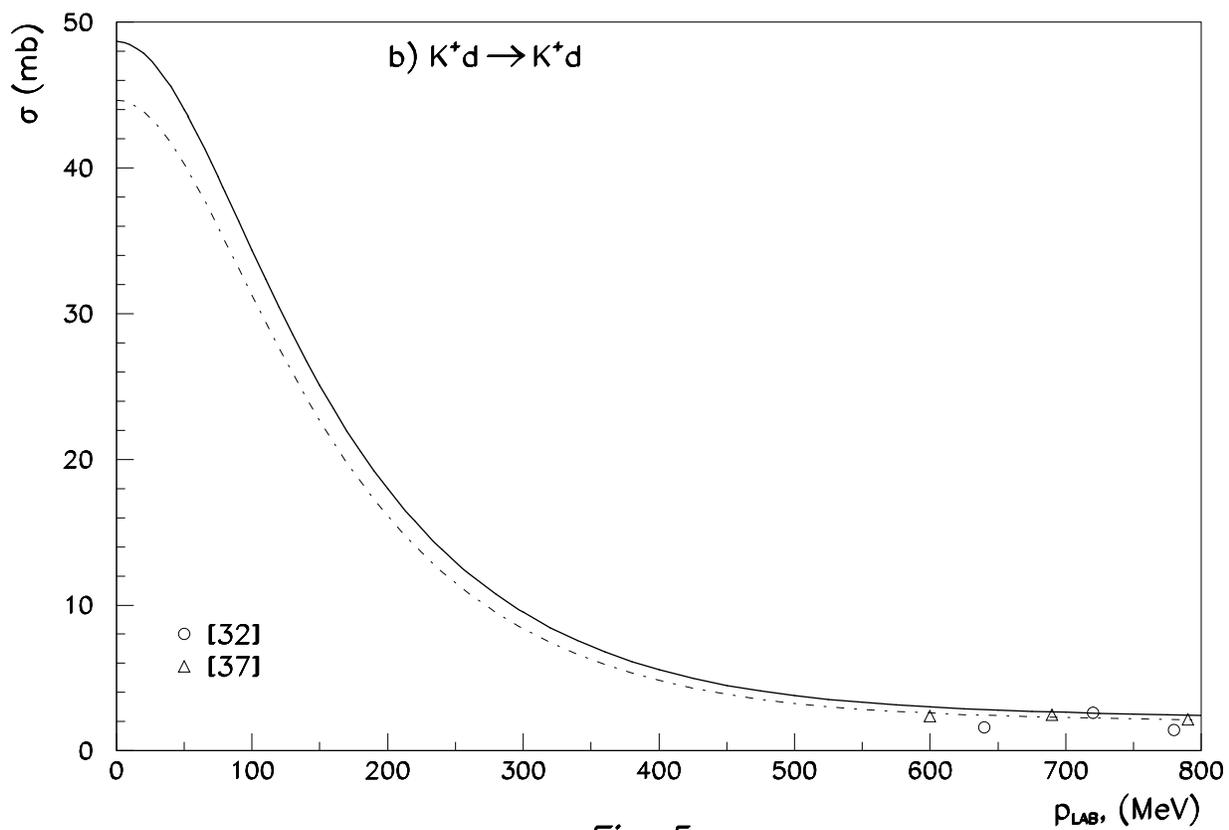
\item[Fig.~5:~]
The $K^+d$ total ($a$) and elastic ($b$) \crss. The symbols correspond
to the experimental data (see references on the plot). The curves show
the results of computations. The total $K^+d$ \crs\ (solid curve in
Fig.~5$a$) is a sum of \crss, shown by solid curves in Figs.~4$a$, 4$b$,
and 5$b$. The dashed curve in Fig.~5$a$ corresponds to the same sum
but with contribution of the $\kdpn$ \crs\ (dashed curve in Fig.~4$a$),
not corrected for NN~FSI. The dashed-dotted curves show the results,
obtained with the $s$-wave part of DWF.

\vspace{2mm}
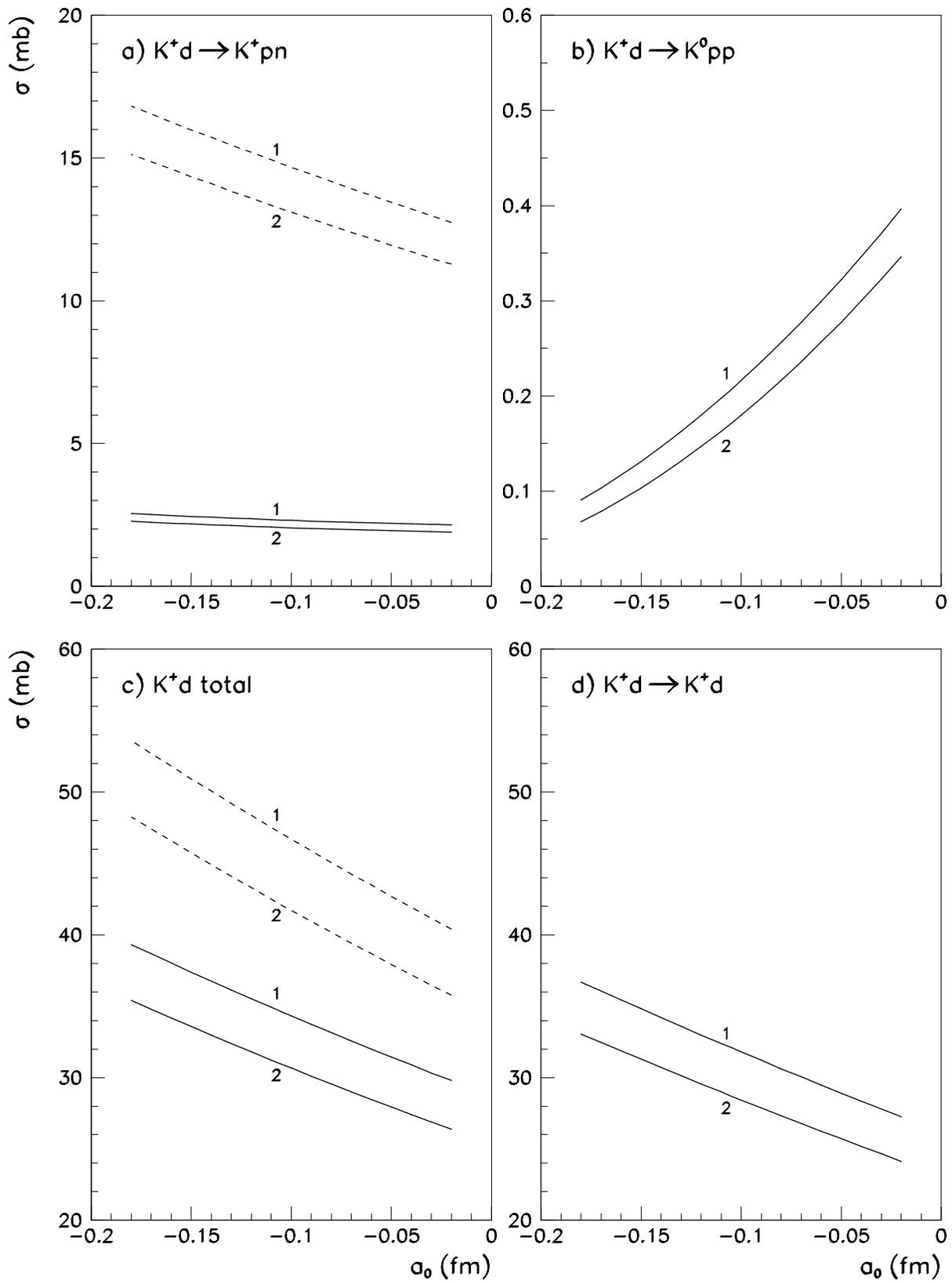
\item[Fig.~6:~]
Predictions for the $K^+d$ \crss\ at the beam momentum
$\PLA =127$~MeV$/c$ as the functions of the isoscalar $KN$ scattering
length $a^{}_0$. The plots show the $\kdpn$ ($a$), $\kdpp$ ($b$),
$K^+d$ total ($c$), and the elastic ($d$) \crss. The curves~1 and 2 show
the results for the values (taken from the Table) $a^{}_1=-0.328~fm$ and
$a^{}_1=-0.308~fm$, respectively. The solid (dashed) curves in
Figs.~$a,c$ show the results, obtained with (without) NN~FSI correction
to the $\kdpn$ \crs.

\end{itemize}


\end{document}